\documentclass[aps, twocolumn, groupedaddress, 10pt]{revtex4-1}

\usepackage{graphicx}
\usepackage{color}
\usepackage{amsfonts, amsmath, amsthm, amscd}

\def\epp{\:.}
\def\epc{\:,}

\def\Isp{I_\text{sp}}
\def\rhosp{\rho_\text{sp}}

\def\limth{\lim\nolimits_\text{th}}
\def\SYY{S_{\text{YY}}}
\def\Drho{\mathcal{D}[\rho]}

\begin{document}
\preprint{APS/123-QED}

\title{Solution for an interaction quench in the Lieb-Liniger Bose gas}

\author{Jacopo De Nardis}
\affiliation{Institute for Theoretical Physics, University of Amsterdam, Science Park 904\\
Postbus 94485, 1090 GL Amsterdam, The Netherlands}

\author{Bram Wouters}
\affiliation{Institute for Theoretical Physics, University of Amsterdam, Science Park 904\\
Postbus 94485, 1090 GL Amsterdam, The Netherlands}

\author{Michael Brockmann}
\affiliation{Institute for Theoretical Physics, University of Amsterdam, Science Park 904\\
Postbus 94485, 1090 GL Amsterdam, The Netherlands}

\author{Jean-S\'{e}bastien Caux}
\affiliation{Institute for Theoretical Physics, University of Amsterdam, Science Park 904\\
Postbus 94485, 1090 GL Amsterdam, The Netherlands}

\date{\today}

\begin{abstract}
We study a quench protocol where the ground state of a free many-particle bosonic theory in one dimension is let unitarily evolve in time under the integrable Lieb-Liniger Hamiltonian of $\delta$-interacting repulsive bosons. 
By using a recently-proposed variational method, we here obtain the exact non-thermal steady-state of the system in the thermodynamic limit, and discuss some of its main physical properties.
Besides being a rare case of a thermodynamically exact solution to a truly interacting quench situation, this interestingly represents an example where a naive implementation of the generalized Gibbs ensemble fails. 
\end{abstract}

\maketitle

\section{Introduction}
Much interest has recently been devoted to improving our understanding of relaxation in isolated many-body quantum systems, fuelled in particular by example realizations using cold atoms \cite{review}. The combination of strong correlations and off-equilibrium initial conditions makes theoretical treatments arduous, in particular in one dimension where quantum fluctuations and nonperturbative effects are inevitable \cite{GiamarchiBOOK}. 
The main focus of recent studies was to consider quenches \cite{rev} in which a Hamiltonian parameter is suddenly changed, and to investigate the late-time asymptotics of the system's properties. Interestingly, although one naively expects that ergodicity generally leads to thermal Gibbs distributions, it has now become clear that non-thermal distributions can sometimes occur, more specifically in circumstances in which nontrivial conservation laws exist. A generalized Gibbs ensemble (GGE) must then be used, in which all conserved charges obtain their own effective temperatures, these being set by initial conditions \cite{GGE}.

Though extremely appealing, the implementation of the GGE poses serious challenges, among which the fact that for many models conservation laws are difficult to handle; calculating the generalized effective temperatures is usually impossible (though exceptional cases where such calculations can be carried through exist \cite{2010_Fioretto_NJP_12,2012_Caux_PRL_109, 2013_Fagotti_JSTAT}). The GGE was thus mostly implemented for theories which are mappable to free systems, for which the momentum occupation modes can be used as conserved charges \cite{rev, GGE, ising}.

In this letter, we deal with a specific quench problem of recent interest \cite{LLquench, 2010_Gritsev_JSTAT_P05012, 2010_Carusotto, 2013_Kormos_arXiv_1305}, namely the interaction quench in the Lieb-Liniger (LL) Bose gas and more specifically the release of the noninteracting Bose-Einstein condensate (BEC) ground state into a system with finite repulsive interactions. Besides being of experimental interest \cite{One_Dim_Bos_Gas,J. J. P. van Es_2010,Wright_2013}, this case surprisingly cannot be treated theoretically using the standard GGE, due to creeping infinities in the expectation values of the conserved charges \cite{2013_Kormos_arXiv_1305}. 

Recently, an alternative approach has been proposed to deal with integrable systems with out-of-equilibrium initial conditions \cite{2013_Caux_PRL_110}. This scheme, which is based on a generalized Thermodynamic Bethe Ansatz (TBA) \cite{2012_Caux_PRL_109,2012_Mossel_JPA_45}, is a thermodynamically exact variational method using as input the overlaps of the initial state with the eigenfunctions of the Hamiltonian driving the post-quench time evolution. We refer to it as the `quench action' approach; it was up to now only tested on free systems for which independent results were available.

Using the quench action, we here provide an exact solution to the BEC-to-LL quench problem.
This allows us in particular to access the physical properties of the non-thermal steady state at long times after the quench,
and represents to our knowledge the first example of a quench to a truly interacting system for which exact results are obtained in the thermodynamic limit. Our results would be applicable to experiments in ring-like geometries \cite{Wright_2013} or box-like potentials \cite{J. J. P. van Es_2010}.
The overall method we use, being quite generic, forms a blueprint for potentially treating many other quench situations.

\section{The quench protocol}
We consider a system of $N$ bosons on a one-dimensional ring of circumference $L$ and impose periodic boundary conditions. Our initial state is the ground state in the absence of interactions, {\it i.\,e.}~the BEC state $|0\rangle$ with $\langle \mathbf{x} | 0 \rangle = \psi_0 (\mathbf{x}) = \frac{1}{L^{N/2}}$.
At $t = 0$, we suddenly turn on interparticle interactions; the time evolution is thus from that moment onwards driven by the Lieb-Liniger Hamiltonian \cite{1963_Lieb_PR_130_1}
(setting $\hbar = 2m = 1$)
\begin{equation}\label{eq:LL_Ham}
H_{LL}=  -\sum_{j=1}^N \frac{\partial^2 }{\partial x^2_j} + 2c \sum_{j>k} \delta(x_j - x_k)
\end{equation}
in which $c$ parametrizes the interaction strength. We here focus on the repulsive regime $c > 0$. The exact eigenstates of \eqref{eq:LL_Ham} are Bethe Ansatz wave functions,
\begin{equation} \label{eq:EF}
\Psi\left({\boldsymbol{x}}|{\boldsymbol{\lambda}}\right) = F_{\boldsymbol{\lambda}} \sum_{P\in \mathcal{S}_N}  A_P (\boldsymbol{x}|{\boldsymbol{\lambda}}) \prod_{j=1}^N   e^{i\lambda_{P_j}x_j} \epc
\end{equation}
with $ F_{\boldsymbol{\lambda}} = \frac{\prod_{j>k=1}^N (\lambda_j -\lambda_k)}{\sqrt{ N!  \prod_{j>k=1}^N\left( (\lambda_j - \lambda_k)^2 + c^2\right)}}$ and $A_P ({\boldsymbol{x}}|{\boldsymbol{\lambda}}) = \prod_{j>k=1}^N \left( 1 - \frac{ic \: \text{sgn}(x_j - x_k)}{\lambda_{P_j} - \lambda_{P_k}}\right)$. Under periodic boundary conditions, rapidities $\boldsymbol{\lambda} \equiv \{\lambda_j\}_{j=1}^N$ get quantized and are required to satisfy the Bethe equations \cite{1963_Lieb_PR_130_1}
\begin{equation}\label{eq:BE_Log}
  \lambda_j =  \frac{2 \pi I_j}{L} - \frac{2}{L} \sum_{k=1}^N \arctan{\left(\frac{\lambda_j - \lambda_k}{c }\right)} \epp
\end{equation}
The set of quantum numbers $I = \{I_j\}_{j=1}^N$, which are mutually distinct integers (half-odd integers) for N odd (even), labels an eigenstate $|I\rangle$ uniquely. Such a state has momentum 
$P_{I} = \sum_{j=1}^N\lambda_j$ and energy $\omega_I = \sum_{j=1}^N \lambda_j^2$. Higher conserved charges $\{ \hat{Q}_n\}_{n\in\mathbb{N}}$ \cite{1990_Davies_PA_167} have eigenvalues $\hat{Q}_n | I\rangle = \sum_{j=1}^N \lambda_j^n |I\rangle$. The norm of the wave function~\eqref{eq:EF} is given by the determinant of the Gaudin matrix \cite{KorepinBOOK}. 

Given such a basis of energy eigenstates, the exact time evolution of a generic normalized initial state $| \psi_0 \rangle $ can be formally written as
\begin{equation}
|\psi_0 (t) \rangle = \sum\nolimits_{I} e^{- S_I - i \omega_I t} |I\rangle \epc
\end{equation}
where we introduced the logarithm $S_I = -\log{\langle I | \psi_0\rangle}$ of the overlap coefficient between a normalized Bethe state and the initial state. The expectation value of a generic operator $\mathcal{O}$ on the initial state at any time $t$ is then
\begin{equation}\label{eq:Double_sum}
\langle \psi_0 | \mathcal{O}(t ) | \psi_0 \rangle =
   \sum\nolimits_{I,I' }    e^{- S_I^\ast - S_{I'} }   e^{i (\omega_{I}  - \omega_{I'}) t}  \langle I | \mathcal{O} | I' \rangle \epp
\end{equation}

\section{The overlaps}\label{par:Ther_overlap}
The first challenge is to compute overlaps between the initial BEC state and the Bethe eigenstates \eqref{eq:EF}. The only states with nonzero overlap with the BEC state are parity-invariant Bethe states such that for each positive rapidity its negative counterpart is also present. Considering $N$ even, we denote such states by $|\lambda,-\lambda\rangle \equiv | \{\lambda_j\}_{j=1}^{N/2} \cup \{-\lambda_j \}_{j=1}^{N/2} \rangle$ where all $\lambda_j$ are taken to be positive. 
The parity invariance is a straightforward consequence of the conservation of momentum and all other odd charges during the quench. 
This can be easily checked by computing matrix elements of the conserved charges $\hat{Q}_{2m+1}$ with respect to the BEC state and the Bethe state, 
\begin{equation}
0 = \langle 0 | \hat{Q}_{2m+1} | I \rangle = \langle 0 |  I \rangle \sum_{j=1}^N \lambda_j^{2m+1} \epp
\end{equation}

The overlap of the initial BEC state with a parity-invariant Bethe state is
\begin{equation}\label{eq:Overlaps}
\langle  \lambda,-\lambda | 0 \rangle = \sqrt{ \frac{(cL)^{-N}N!}{ \det_{j,k=1}^{N}G_{jk}  } }  \frac{\det_{j,k=1}^{N/2} G^{Q}_{jk}} {{\displaystyle \prod\limits_{j=1}^{N/2} \frac{\lambda_j}{c}   \sqrt{\frac{\lambda_j^2}{c^2} + \frac{1}{4} } } }\epp
\end{equation}
The matrix $G_{jk}^{Q}$ is of the same form as the Gaudin matrix $G_{jk}$, but with a different kernel:
\begin{equation}\label{eq:Gaudin_Q}
G^Q_{jk} = \delta_{jk} \Big( L + \sum_{l=1}^{N/2}\, K^Q(\lambda_{j},\lambda_{l}) \Big) - K^Q (\lambda_{j},\lambda_{k}) \epc
\end{equation}
where $K^{Q}(\lambda,\mu) = K(\lambda - \mu)+ K(\lambda + \mu)$, with $K(\lambda) = 2c/(\lambda^2 + c^2)$. Expression (\ref{eq:Overlaps}) was analytically verified up to $N=8$ and then proven for arbitrary $N$ in Ref.~\cite{2014_Proof_Micheal}.

\section{Quench action approach}
The next step is to evaluate Eq.~\eqref{eq:Double_sum}. The difficulty represented by the double Hilbert-space sum is substantial, and we follow the approach proposed in Ref.~\cite{2013_Caux_PRL_110} to handle it (see also Ref.~\cite{QuenchActionPreprint} for more details). In the thermodynamic limit $L \to \infty$
with fixed density $n=N/L$ 
(which we denote as $\limth$) a single sum over the Hilbert space is replaced by a functional integral over positive smooth functions $\rho(\lambda)$, each function describing the density of Bethe roots for an ensemble of states with Yang-Yang entropy \cite{1969_Yang_JMP_10} 
\begin{equation}
S_{YY}[\rho]= L \int_{-\infty}^{\infty} d\lambda \big( (\rho +  \rho^h) \ln (\rho + \rho^h) -   \rho \ln \rho
  - \rho^h \ln \rho^h   \big) \epp
\end{equation}  
The hole density 
$\rho^h$ is related to the particle density $\rho$ by the thermodynamic form of the Bethe equations
\begin{equation}\label{eq:BE_ThLim}
\rho(\lambda) + \rho^h
(\lambda) = \frac{1}{2\pi} + \int_{-\infty}^{\infty} \frac{d\mu}{2\pi} K(\lambda - \mu) \rho(\mu)\epp
\end{equation}

Explicitly, when dealing with a quantity ${\cal O}_I$ which scales to a smooth functional ${\mathcal O}[\rho]$ in the thermodynamic limit, we can write
\begin{equation}
\limth \sum\nolimits_{I} {\cal O}_I = \int \Drho \:   e^{\SYY[\rho]} {\cal O}[\rho] \epc
\end{equation}
up to an overall normalization constant. Focusing on generic operators with negligible matrix elements between states that scale to different distributions $\rho$, expression \eqref{eq:Double_sum} transforms to
\begin{align} \label{eq:QA_expectation}
&\langle \psi_0 | \mathcal{O}(t ) | \psi_0 \rangle = \frac{1}{2} \int \Drho \:   e^{ -2S[\rho] + \SYY[\rho]} \times\notag\\ 
&\sum_{ \mathbf{e} } \Big(   e^{ - \delta s_\mathbf{e} -  i \delta\omega_\mathbf{e} t } \langle \rho,\emptyset | \mathcal{O} | \rho , \mathbf{e} \rangle +   e^{ - \delta s^*_\mathbf{e} +  i \delta\omega_\mathbf{e} t } \langle \rho,\mathbf{e} | \mathcal{O} | \rho , \emptyset \rangle \Big) \epc 
\end{align}
involving a sum over the set of discrete particle-hole excitations on $\rho$ denoted by $\mathbf{e} = \{p_i, h_i \}_{i=1}^m$, $m= 0,1,\ldots \epc$ with
energy $\delta\omega_\mathbf{e}$. The extensive real part of the overlap coefficient is denoted by
\begin{equation}
S[\rho]  =   \limth \Re S_I \epc
\end{equation}
and we used $\delta s_\mathbf{e}$ to denote the relative overlaps of states that are equal up to a set $\mathbf{e}$ of particle-hole excitations, 
\begin{equation}
\delta s_{\mathbf{e}} = -\limth \log \big(  \langle  I \cup \mathbf{e} | \psi_0 \rangle / \langle I | \psi_0 \rangle \big) \epp
\end{equation}
The extensive quench action $S^Q [\rho]  \equiv 2S[\rho] - \SYY[\rho]$ is real and bounded from below which guarantees the convergence of the functional integral. In the thermodynamic limit this can be 
exactly evaluated using the saddle point of the quench action, which is fixed by the condition
\begin{equation}\label{eq:Saddle_point}
\left.\frac{\delta S^Q[\rho] }{\delta\rho} \right|_{\rhosp} = \left.\frac{\delta ( 2S[\rho] - \SYY[\rho])}{ \delta \rho }\right|_{\rhosp} = 0 \epp
\end{equation}
An explicit solution can then be obtained using the generalized Thermodynamic Bethe Ansatz \cite{2012_Mossel_JPA_45,2012_Caux_PRL_109}.

Putting everything together leads to a much simpler expression for the full time evolution \eqref{eq:Double_sum} in terms of matrix elements of states around the saddle point distribution:
\begin{multline}\label{eq:QA_expectation_2}
\limth\langle \psi_0 | \mathcal{O}(t ) | \psi_0 \rangle =
\frac{1}{2} \sum_{ \mathbf{e} }   e^{ - \delta s_\mathbf{e} -  i \delta\omega_\mathbf{e} t } \langle \rhosp,\emptyset | \mathcal{O} | \rhosp , \mathbf{e} \rangle \\
+ \frac{1}{2} \sum_{ \mathbf{e} }    e^{ - \delta s^\ast_\mathbf{e} +  i \delta\omega_\mathbf{e} t } \langle \rhosp,\mathbf{e} | \mathcal{O} | \rhosp , \emptyset \rangle \epp 
\end{multline}
Expression \eqref{eq:QA_expectation_2} is exact and valid for any time $t$ after the quench. In particular it recovers the expectation value in the infinite time limit where no time average is involved
\begin{equation}
\lim_{t \to \infty} \limth\langle \psi_0 | \mathcal{O}(t ) | \psi_0 \rangle=  \langle \rhosp |   \mathcal{O} | \rhosp \rangle \epp
\end{equation}
As discussed in Ref.~\cite{2013_Caux_PRL_110}, this can be viewed as a generalization of the Eigenstate Thermalization Hypothesis \cite{ETH} similar to that proposed in Ref.~\cite{2011_Cassidy_PRL_106}. It also relates to the time-averaged approach of Ref.~\cite{2013_Mussardo_arXiv_1308}.

\section{Explicit solution to the saddle point equation}
With the previously computed overlaps, we are now in position to apply the quench action approach. We need only their extensive part $S[\rho]$, which can be extracted from \eqref{eq:Overlaps} by taking the thermodynamic limit (see Appendix \ref{app_over})
\begin{multline}
S[\rho] = -\limth \Big(\log \langle \lambda,-\lambda | 0 \rangle\Big) = \frac{L n}{2 } \left(\log \gamma +1\right) \\
 +\frac{L}{2} \int_0^\infty d\lambda\rho(\lambda) \log{\left[ \frac{\lambda^2}{c^2} \left(\frac{\lambda^2}{c^2} + \frac{1}{4} \right)\right]} + \mathcal{O}(L^0)  \epc 
\end{multline}
where $\gamma = c/n$. We only need the $\rho$-dependent part of the overlap $S[\rho]$ which acts as a `driving term' in the generalized TBA equation \eqref{eq:Saddle_point}. The quench action then reads
\begin{multline}
S^Q[\rho]/L = \int_0^\infty d\lambda\Big[ \rho(\lambda) \log\left(\frac{\lambda^2}{c^2} \left( \frac{1}{4} + \frac{ \lambda^2}{c^2}  \right) \right) \\
 -   \rho^t(\lambda) \log\rho^t(\lambda) + \rho(\lambda) \log\rho(\lambda)  + \rho^h(\lambda) \log\rho^h(\lambda) \Big] \epp
\end{multline}
Note that the Yang-Yang entropy has non-zero measure only on the half space $\lambda>0$ since the filling of quantum numbers associated with the negative rapidities is unambiguously determined by the positive ones and thus do not contribute to the entropy of the state. To impose the normalization condition on $\rho$ we add a Lagrange multiplier $h$ as in Ref.~\cite{KorepinBOOK} which can be viewed, in the spirit of the free energy, as a generalized chemical potential. This corresponds to modifying our functional measure as
\begin{multline}
\int \Drho\:   e^{- S^Q[\rho]}  \to  \\ 
\int_{- i \infty}^{+i \infty}dh\int \Drho\:   e^{- S^Q[\rho]} 
e^{ -\frac{L h}{2} \left(n-\int_{-\infty}^\infty d\lambda \rho(\lambda) \right)   }\epp
\end{multline}
Taking the functional derivative with respect to $\rho$ 
and using dimensionless quantities $x=\lambda/c$ and $K(x)=\frac{2}{x^2+1}$, we obtain as a saddle point equation a non-linear integral equation for the function $a(x) \equiv \rho(\lambda)/\rho^h(\lambda)$, 
\begin{multline}\label{eq:NLIE}
	\log{a(x)} = \log{(\tau^2)} - \log{\left[x^2\left(x^2+1/4\right)\right]} \\
	+\int_{-\infty}^{\infty} K(x-y) \log{\left[1+a(y)\right]}   \frac{dy}{2\pi}\epc
\end{multline}
where $\tau$ is related to the Lagrange multiplier $h$ via $\tau=  e^{h/2}$. 
The functions $a$ and $\rho$ are directly connected. Taking the derivative $\frac{\tau}{2}\partial_\tau$ of Eq.~\eqref{eq:NLIE} and due to the thermodynamic form of the Bethe equations \eqref{eq:BE_ThLim} the function $2\pi\rho(x)$ is given by $\frac{\tau}{2}\partial_\tau a(x)/(1+a(x))$. 

{The non-linear integral equation \eqref{eq:NLIE} has an analytical solution, which can be derived as follows.} In the limit $\tau\to 0$ the driving term becomes large and negative for all fixed $x > \tau$ and the convolution integral gives only subleading contributions. Thus, the first non-trivial order of the function $a(x)$ reads
\begin{equation}
	a^{(0)}(x)=\frac{\tau^2}{x^2(x^2+1/4)}\epp
\end{equation}
In order to calculate the next leading term, we plug this result into the convolution integral on the right hand side of the saddle point equation \eqref{eq:NLIE}.  Using the relation
\begin{equation}\label{eq:int_log_formula_SupMat}
	\int_{-\infty}^{\infty}\frac{1/\pi}{(x-y)^2+1}\log{\left[y^2+\alpha^2\right]}  d  y = \log{\left[x^2+(\alpha+1)^2\right]}\epc 
\end{equation}
we obtain up to order $\tau^2$
\begin{multline}
	\log{a}(x) = \log{(\tau^2)} \\-  \log{\left[x^2\left(x^2+1/4\right)\left(x^2+1\right)\left(x^2+9/4\right)\right]}
	\\
	+\int_{-\infty}^{\infty} K(x-y) \log{\left[y^2(y^2+1/4)+\tau^2\right]}   \frac{dy}{2\pi} \epp
\end{multline}
By rewriting $y^2(y^2+1/4)+\tau^2=(y^2+y_{-}^2)(y^2+y_{+}^2)$ where $y_\pm=\frac{1}{\sqrt{8}}\sqrt{1\pm\sqrt{1-64\tau^2}}$, using Eq.~\eqref{eq:int_log_formula_SupMat} again and expanding $y_\pm$ to lowest order in $\tau$: $y_{+}=1/2$ and $y_{-}=2\tau$, we get
\begin{multline}	a(x)=\frac{\tau^2\left(x^2+(1+y_{-})^2\right)\left(x^2+(1+y_{+})^2\right)}{x^2(x^2+1/4)(x^2+1)(x^2+9/4)} \\
	=\frac{\tau^2}{x^2(x^2+1/4)} \left[1+\frac{4\tau(1+\tau)}{x^2+1}\right]\epp
\end{multline}
Hence, the function $a(x)/a^{(0)}(x)$ up to first order in $\tau$ reads 
\begin{equation}
	\frac{a^{(1)}(x)}{a^{(0)}(x)} = 1+\frac{4\tau}{x^2+1}\epp
\end{equation}
Repeating this procedure we can calculate higher orders in $\tau$ systematically, leading to an expression up to generic order $\tau^N$
\begin{equation}
\frac{a^{(N)}(x)}{a^{(0)}(x)}= \sum_{n=1}^{N+1} \binom{2n}{n-1} \prod_{j=2}^n \frac{\tau}{x^2 + (j/2)^2}\epp
\end{equation}
The limit $N\to\infty$ leads to the solution of the saddle point equation \eqref{eq:NLIE}, 
\begin{align}
a(x) &= \lim\limits_{N\to\infty}\,a^{(N)}(x) = \sum_{n=1}^\infty \binom{2n}{n-1}  \prod_{j=0}^n \frac{\tau}{x^2 + (j/2)^2} \notag\\[1ex]
&= \frac{2\pi\tau}{x\sinh{(2\pi x)}} I_{1-2 i x}\big(4\sqrt{\tau}\big)I_{1+2 i x}\big(4\sqrt{\tau}\big)\epp \label{eq:exacta}
\end{align}
Here, $I_n(z)$ is the modified Bessel function of the first kind of order $n$. We eventually get
\begin{equation}\label{eq:rho}
2 \pi\rho(x) = \frac{\frac{\tau}{2}\partial_\tau a(x)}{1+a(x)} = \frac{a(x)}{1+a(x)} \frac{\tau\partial_\tau }{2}\log a(x) \epp
\end{equation}
Note that the saddle point distribution $\rhosp$ is then given by $\rhosp(\lambda) = \rho(\lambda/c)$.
 
The computation of the dimensionless particle density $N/(Lc) = n/c=1/\gamma$ and energy density $E/(Lc^3) = e/c^3$ by numerical integration of function \eqref{eq:rho} yields
\begin{align}
	\frac{n}{c} &= \int_{-\infty}^\infty \rho(x)  d  x = \tau\epc \qquad \frac{e}{c^3} = \int_{-\infty}^\infty x^2 \rho(x)  d  x = \tau^2\epc
\end{align}
{\it i.e.}~$\tau=1/\gamma$ and $e=cn^2=\gamma n^3$, in agreement with the initial energy density
\begin{equation}
 \limth L^{-1}\langle 0 | H_{LL} | 0 \rangle = \gamma n^3 \epp
\end{equation}

\begin{figure}
\begin{center}
\includegraphics[width=0.95\columnwidth]{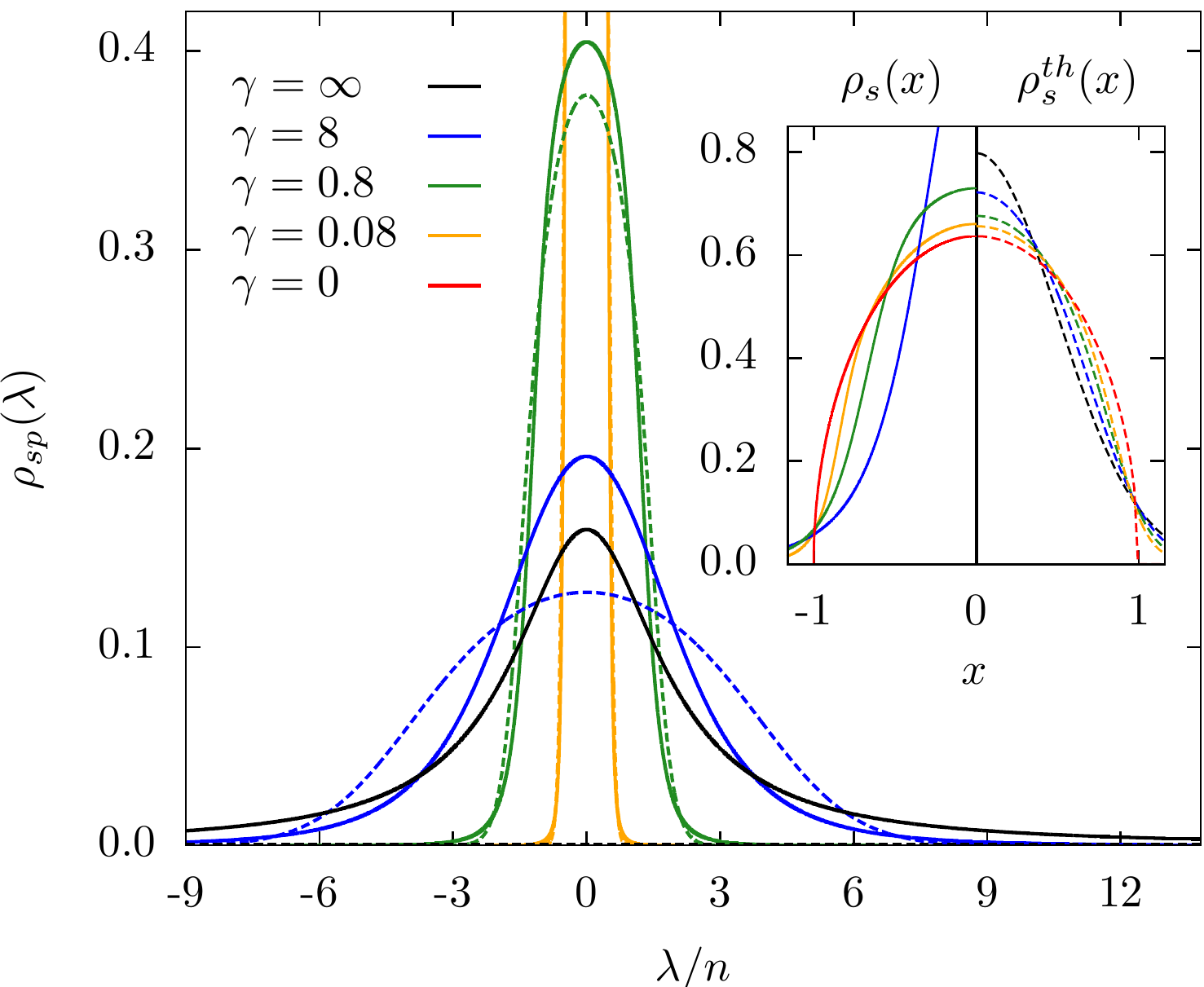}
\caption{\label{fig:rhos} (color online) Density function $\rhosp(\lambda)$ in \eqref{eq:rho} for different $\gamma$'s for the saddle point state (solid lines) and for the thermal state (dashed lines). In the main plot (resp. inset), $\rhosp(0)$ (resp.  $\rho_{th}(0)$) is a decreasing (increasing) function of $\gamma$.
\textit{Inset}: Scaled functions $\rho_s(x)=2\sqrt{\gamma}\rho(2n\sqrt{\gamma}x)$ and $\rho_s^{\text{th}}(x)$ which approach the semi-circle $\frac{2}{\pi}\sqrt{1-x^2}$ in the limit $\gamma\to 0$.}
\end{center}
\end{figure}

\section{Physical properties of the steady state.}
\begin{figure}\begin{center}
\includegraphics[width=0.95\columnwidth]{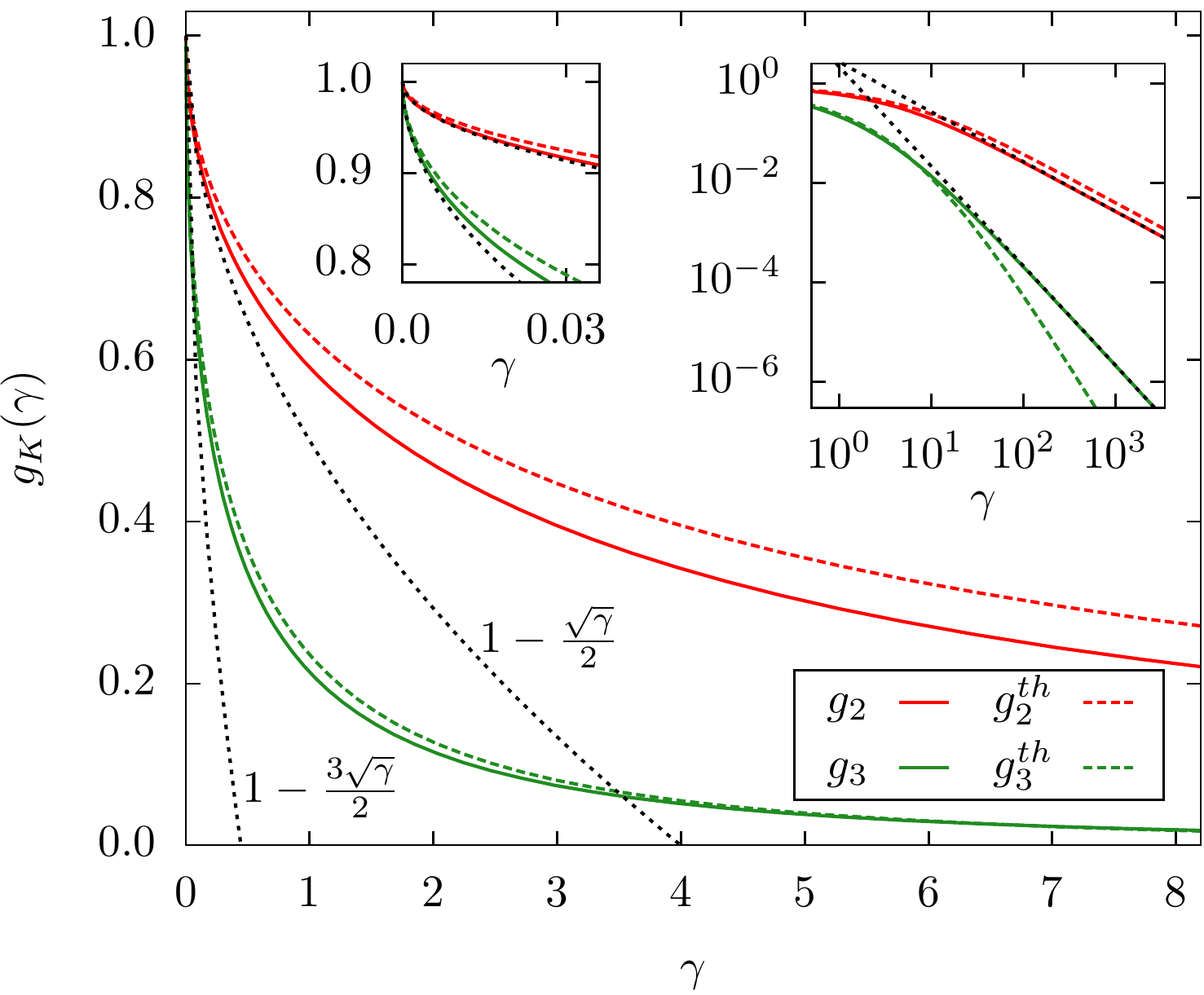}
\caption{\label{fig:g2g3} (color online) Expectation values $g_2$ (red, upper curve) and $g_3$ (green, lower curve) as function of $\gamma$ on the exact saddle point state (solid lines) and on the thermal one (dashed lines). Asymptotic behaviours (black dashed lines) $g_2 \sim 8/(3\gamma)$, $g_3 \sim  32/(15\gamma^2)$ for $\gamma \to \infty$ as in \cite{2013_Kormos_arXiv_1305} and $g_2 \sim 1-\sqrt{\gamma} /2$, $g_3 \sim  1- 3\sqrt{\gamma}/2$ for $\gamma \to 0$. \textit{Insets}: Same plot (in logarithmic scale) for different ranges of $\gamma$.} 
\end{center}\end{figure}

The saddle point distribution gives us access, in principle, to any correlation function of operators allowed within the quench action approach. We focus on density correlations. Following the method of \cite{2011_Kormos_PRL_107,2011_Pozsgay_JSTAT_P01011} we compute in Fig.~\ref{fig:g2g3} the static density moments $g_2$ and $g_3$ defined by 
\begin{equation}
g_K = \langle \rhosp | :\!(\hat{\rho}(0)/n)^K\!\!: | \rhosp \rangle\epp
\end{equation}
The density operator is defined as $\hat{\rho}(x) =\mathbf{\Psi}^{\dagger}(x)\mathbf{\Psi}(x)$, where the operators $\mathbf{\Psi}(x)$ satisfies the canonical commutation relations $[\mathbf{\Psi}(x), \mathbf{\Psi}^{\dagger} (x')]= \delta(x-x')$. Results for the saddle point state are compared with the ones obtained using a thermal state at fixed particle density $n$ and energy density $e = n^3 \gamma$. These results clearly display the lack of thermalization long after the quench.

\begin{figure}\begin{center}
\includegraphics[width=0.95\columnwidth]{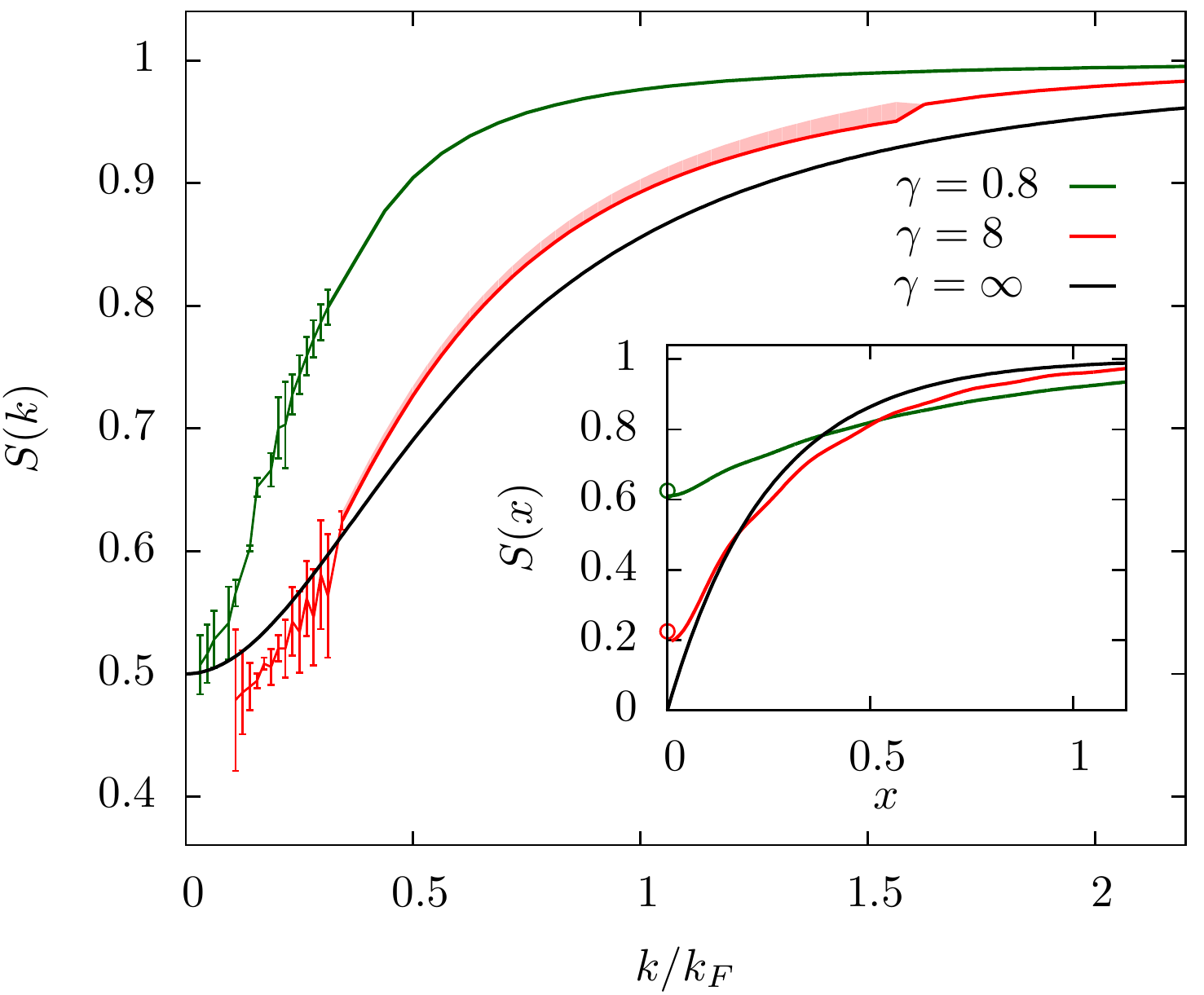}
\caption{\label{fig:S_x_S_k} (color online) Density-Density correlation on the saddle point state in momentum space $k$ (in unit of $k_F = \pi n$) and (\textit{inset}) in real space $x \in [0,L]$ (from top to bottom for increasing values of $\gamma$). At $x \to 0$ the numerical results approach the analytical ones for $g_2$ (open dots). Curves are obtained by joining datas from system sizes $N=64$ (small $k$), $32$ and $8$ (large $k$). Error bars and shaded region are respectively estimates of finite-size discretization errors or missing intensity based on sumrule saturation levels. 
At $k=0$ numerical datas suggest $S(k) \to 1/2$ irrespective of the interaction, which agrees with the Bogoliubov prediction for small $\gamma$ \cite{2010_Carusotto}. }
\end{center}\end{figure}

In Fig.~\ref{fig:S_x_S_k} we further address static correlations by studying the static structure factor $S(x) = \langle \Isp| \hat{\rho}(x) \hat{\rho}(0) | \Isp \rangle$ and its Fourier transform using the Lehmann representation
\begin{equation}
	S(k) = L \sum_{I} | \langle \Isp | \hat{\rho}(0) | I \rangle |^2  \delta_{k,P_I} \epc
\end{equation}
where the state $|\Isp\rangle$ scales to the saddle point distribution. The matrix elements are known exactly through the method of Algebraic Bethe Ansatz \cite{1990_Slavnov_TMP_79_82} and summed into correlations by the ABACUS algorithm \cite{2009_Caux_JMP_50} following the method in Ref.~\cite{Panfil_in_prep}.

\section{Time Evolution towards the steady state} 
Within the quench action logic, the full time evolution is recovered by excitations around the saddle point state according to Eq.~\eqref{eq:QA_expectation_2}. For the special case of a quench to the Tonks-Girardeau gas the density operator creates only single particle-hole excitations, and the complete time evolution of the density-density correlation can be calculated explicitly by simply summing over all parity-invariant pairs of particle-hole excitations {(see Appendix \ref{sec:app_time_evo})}. We then obtain
\begin{multline}
\langle \hat{\rho}(x) \hat{\rho}(0) \rangle_t 
- \langle\rhosp|\hat{\rho}(x) \hat{\rho}(0)|\rhosp\rangle
\\=\left|\int_{-\infty}^\infty \frac{dk}{ \pi}  \frac{n k}{4 n^2 + k^2 }   e^{- 2it k^2  + i k x }  \right|^2
\end{multline}
where $\langle\rhosp|\hat{\rho}(x) \hat{\rho}(0)|\rhosp\rangle = n\delta(x) + n^2(1 -   e^{-4n|x|})$. This easily reproduces the recently-calculated result of Ref.~\cite{2013_Kormos_1307} and therefore represents a nontrivial check of the validity of expression \eqref{eq:QA_expectation_2} for any time $t$ after the quench.

\section{Conclusion and outlook}
In this paper the logic of the quench action was used to obtain an exact description of the quench from the non-interacting Lieb-Liniger bose gas in its ground state to arbitrary repulsive interaction described by the parameter $c>0$. A generalized TBA equation was derived from an exact expression for the overlaps and subsequently analytically solved. For late times after the quench the system is well described by a saddle-point state that is very different from any thermal state, as the distribution of rapidities has a polynomial $\lambda^{-4} $ tail for any nonzero value of the final interaction $c$. This tail leads to the divergence of all conserved charges $Q_{2n}$ for $n>1$ (the divergence of $Q_4$ indicating infinite energy fluctuations) and thus the inapplicability of the standard GGE logic, since the chemical potentials associated with the conserved charges cannot be determined from expectation values on the initial state. The GGE free energy would be of the form $\sum_n \beta_n Q_n$ with $Q_n = \sum_j \lambda_j^n$. The overlaps however give terms of the form $\log \lambda$ in the quench action which have no expansion in powers of $\lambda$. It should be noted though that in the limit of small post-quench interaction parameter the thermal and exact distributions become increasingly alike.

The logic also recovers the exact time evolution of expectation values of generic operators at any time after the quench, thereby circumventing the problematic double sum over the full Hilbert space. As a nontrivial confirmation of our method the density-density correlations in the Tonks-Girardeau regime were worked out and verified to give correct results. In future publications, we will investigate the time evolution of observables such as $g_2,g_3$ and the dynamical structure factor $S(k,\omega)$ for any value of the final interaction. {The question of wether the overlap formula \eqref{eq:Overlaps} can be generalized to interacting initial states will also be addressed}. Our method, combined with recent developments in the computation of overlaps in lattice systems \cite{2010_Mossel_overlap, 2013_Balasz_overlap,2014_Brockmann,2014_Proof_Micheal}, opens the way to studying non-equilibrium quench dynamics in other interacting integrable systems such as quantum spin chains. 



\section*{Acknowledgments} We acknowledge useful discussions with M. Panfil, M. Kormos, A. Shashi, Y.-Z. Chou, J. Mossel, M. Rigol, F. H. L. Essler, P. Calabrese, G. Mussardo and I. Carusotto. 
This work is dedicated to the memory of Adilet Imambekov.
We acknowledge support from the Foundation for Fundamental Research on Matter (FOM) and the Netherlands Organisation for Scientific Research (NWO).

\appendix

\section{Thermodynamic limit of the overlaps}\label{app_over}
Let us consider the expression for the overlaps at finite size 
\begin{equation}\label{eq:overlapSupMat}
 \langle \lambda, - \lambda | 0 \rangle   
= \sqrt{ \frac{ {(cL)^{-N}  N !}}{ \det_{j,k=1}^{N}G_{jk}  } }  \frac{\det_{j,k=1}^{N/2} G^{Q}_{jk}}{\prod\limits_{j=1}^{N/2}\frac{ \lambda_j}{c} \sqrt{ \frac{\lambda_j^2}{c^2} + \frac{1}{4}} } \epp
\end{equation}
We want to compute this expression for a generic Bethe state in the thermodynamic limit up to corrections $\mathcal{O}(1/L)$. 

The thermodynamic limit consists in recasting any sum over rapidities of the state as an integral weighted by the density of rapidities. Given a state specified by its set of rapidities $\boldsymbol{\lambda}$ this means that for any smooth function $f$ of the quantum numbers
 \begin{equation}\label{eq:From_sum_To_int}
 \limth \sum_{j=1}^Nf(\lambda_j) = L \int_{-\infty}^\infty d\lambda  \: \rho(\lambda)f(\lambda) + \mathcal{O}(L^0) \epp
 \end{equation}
The subleading corrections depend on how we define microscopically the $\rho(\lambda)$. For each smooth distribution $\rho$ we can choose a ``maximally flat'' state $ | \lambda^{\rho} , - \lambda^{\rho}  \rangle $ such that all $\mathcal{O}(L^0)$ corrections in \eqref{eq:From_sum_To_int} are zero
\begin{equation}
\limth \sum_{j=1}^N f(\lambda_j^\rho) = L \int_{-\infty}^{\infty} d\lambda \: \rho(\lambda) f(\lambda) + \mathcal{O}(1/L) \epp
\end{equation}
By adding a set of $m$ particle-hole excitations to the maximally flat state such that $ \limth (m/N) = 0$ we can split the set of rapidities in ``non-excited ones" $\{ \tilde{\lambda}_j\}_{j=1}^{N-m}$ and in a set of particle excitations  $\{\tilde{\lambda}_j^p \}_{j=1}^m$. The set of holes $\{  \tilde{\lambda}_j^h\}_{j=1}^m$ contains fictitious rapidities which represent the empty slots left by the particle excitations. The rapidities in the first set are related to the ones of the maximally flat state by the shift function \cite{KorepinBOOK}
\begin{equation}
\tilde{\lambda}_j =  {\lambda}_j^\rho + \sum_{k=1}^m \frac{F(\lambda_j| \tilde{\lambda}^p_k ,\tilde{\lambda}^h_k )}{L} + \mathcal{O}(1/L^2) \epc
\end{equation}
where the shift function for a particle-hole excitation is defined by the integral equation
\begin{multline}\label{eq:shift_F}
2 \pi F(\lambda| \mu^p, \mu^h) - \int_{- \infty}^\infty d\mu \: \vartheta(\mu) K(\lambda - \mu) F(\mu| \mu^p, \mu^h) \\
 = \phi(\lambda - \mu^p )+ \phi(\lambda - \mu^h )\epc
\end{multline}
with the scattering matrix given by $\phi(\lambda) = 2 \arctan(\lambda/c)$ and the weight function by $\vartheta(\lambda) = \frac{\rho(\lambda)}{\rho(\lambda) + \rho^h(\lambda)}$. We can thus write
\begin{align}
&\log  \left(\prod_{j=1}^{N/2} \tilde{\lambda}_j/c   \sqrt{  ( \tilde{\lambda}_j/c )^2 + 1/4} \right)  \nonumber \\
= \:\:  & L \int_0^\infty d\lambda \: \rho(\lambda) \log \left(\lambda/c   \sqrt{  ( \lambda/c)^2 + 1/4} \right)  \nonumber \\  
+&  \sum_{k=1}^m \Big[ \int_0^\infty d\lambda \: \rho(\lambda)  \frac{1 + 8 \frac{\lambda^{2}}{c^{2}}}{\lambda \left(1+ 4\frac{\lambda^{2}}{c^{2}}\right)}  F(\lambda | \tilde{\lambda}^p_k , \tilde{\lambda}^h_k ) \nonumber 
\\+ &
 \log \left(\tilde{\lambda}^p_k/c   \sqrt{ ( \tilde{\lambda}^p_k/c)^2 + 1/4}\right)   \nonumber \\
-&
 \log \left(\tilde{\lambda}_k^h/c   \sqrt{  ( \tilde{\lambda}_k^h/c)^2 + 1/4})  \right) \Big] + \mathcal{O}(1/L) \epp
\end{align}
 
Regarding the two determinants in Eq.~\eqref{eq:overlapSupMat}, we can rewrite the determinant of the Gaudin matrix $G_{jk}$ as
\begin{multline}
\det\nolimits_{j,k=1}^{N}G_{jk} = L^N \prod_{j=1}^N\Big( 1 + \frac{1}{L} \sum_{l=1}^N K(\lambda_j - \lambda_l) \Big) \\ \times
 \det\nolimits_{j,k=1}^N \left(\delta_{jk} - \frac{K(\lambda_j - \lambda_k)}{L + \sum_{l=1}^N K(\lambda_k - \lambda_l) } \right) \epp \label{eq:Gaudin_det}
\end{multline}
From the Bethe equations in the thermodynamic limit, Eq.~\eqref{eq:BE_ThLim}, we have
\begin{equation}
1 + \frac{1}{ L} \sum_{l=1}^N K(\lambda_j - \lambda_l)  = 2\pi\,  \rho^t(\lambda_j) + \mathcal{O}(1/L) \epc
\end{equation}
where $\rho^t = \rho + \rho^h$. Furthermore, in the thermodynamic limit the matrix on the right-hand side of Eq.~\eqref{eq:Gaudin_det} becomes an integral operator on the real line:
\begin{multline}
\delta_{jk} - \frac{K(\lambda_j - \lambda_k)}{L + \sum_{l=1}^N K(\lambda_k - \lambda_l) } \quad \to \quad 1 - \frac{\hat{K_{\rho}}}{2\pi} \epc \qquad 
\\
\text{where} \quad  \left(\hat{K_{\rho}}\, g \right) (\lambda) = \int_{-\infty}^{\infty} \mathrm{d}\mu \, K(\lambda - \mu) \frac{\rho(\mu)}{\rho^{t}(\mu)} g(\mu) \epp
\end{multline}
The matrix $G_{jk}^{Q}$ can be analyzed analogously, using instead of $\hat{K}_{\rho}$ the operator $\hat{K}_{\rho}^{Q}$ acting as
\begin{multline}
 \left(\hat{K_{\rho}}\, g \right) (\lambda) = \Theta(\lambda) \\
  \times\int_{0}^{\infty} \mathrm{d}\mu \, \left[K(\lambda - \mu) + K(\lambda + \mu) \right] \frac{\rho(\mu)}{\rho^{t}(\mu)} g(\mu) \epc
\end{multline}
where $\Theta(\lambda)$ is the Heaviside step function. Putting everything together, one finds that in the thermodynamic limit the ratio of determinants in Eq.~\eqref{eq:overlapSupMat} becomes the ratio of two Fredholm determinants \cite{Bornemann_2008} 
\begin{equation}
\limth \left( \frac{\det_{j,k=1}^{N/2}G^Q_{jk}}{ \sqrt{\det_{j,k=1}^{N}G_{jk}}} \right) \ =\  \frac{Det(1 - \frac{\hat{K}_\rho^Q}{2 \pi })}{ \sqrt{Det(1- \frac{\hat{K}_\rho}{2 \pi})}} + \mathcal{O}(1/L)\epp
\end{equation}
Finally we can write the logarithm of the overlap of the BEC state $|0\rangle$ with a generic state $| \tilde{\lambda},- \tilde{\lambda}\rangle$, which scales to the distribution $\rho$ in the thermodynamic limit, as  
\begin{align}
  \limth & \log\left( \langle  \tilde{\lambda}, - \tilde{\lambda} | 0 \rangle \right) \nonumber \\
  = - &
\frac{L}{2} \int_0^\infty d\lambda \: \rho(\lambda)\log{\left[ \frac{\lambda^2}{c^2} \left(\frac{\lambda^2}{c^2} + \frac{1}{4} \right)\right]}  \nonumber \\
- &   \frac{L n}{2} \left(\log\frac{c}{n} + 1 \right) 
+\log\left[ \frac{Det(1 - \frac{\hat{K}_\rho^Q}{2 \pi })}{ \sqrt{Det(1- \frac{\hat{K}_\rho}{2 \pi})}}\right]  \nonumber
\\- &    \sum_{k=1}^m \Big[ \int_0^\infty d\lambda \: \rho(\lambda)  \frac{1 + 8 \frac{\lambda^{2}}{c^{2}}}{\lambda \left(1+ 4\frac{\lambda^{2}}{c^{2}}\right)} F(\lambda | \tilde{\lambda}^p_k , \tilde{\lambda}^h_k ) 
 \nonumber \\
 + &  \log \left(\frac{\tilde{\lambda}^p_k  \sqrt{ ( \tilde{\lambda}^p_k/c)^2 + 1/4}}{\tilde{\lambda}_k^h \sqrt{  ( \tilde{\lambda}_k^h/c)^2 + 1/4}) }\right) \Big]  \epp
\end{align}
We can split this in an extensive part which depends on the distribution $\rho$,  
\begin{multline}
S[\rho] = \frac{L}{2} \Big[    \int_0^\infty d\lambda \: \rho(\lambda)\log{\left[ \frac{\lambda^2}{c^2} \left(\frac{\lambda^2}{c^2} + \frac{1}{4} \right)\right]} \\
+  n \left(\log\frac{c}{n} + 1 \right)\Big] \epc
\end{multline}
and in a non-extensive part $\delta s_{\mathbf{e}}[\rho]$ which does depend on details of the excitations of the state. The latter contains information about the time evolution after the quench and is given by
\begin{align}
\delta s_{\mathbf{e}}[\rho] =\ & -\log\left[ \frac{Det(1 - \frac{\hat{K}_\rho^Q}{2 \pi })}{ \sqrt{Det(1- \frac{\hat{K}_\rho}{2 \pi})}} \right] \nonumber \\ &+  \sum_{k=1}^m 
\Big[ \int_0^\infty d\lambda \: \rho(\lambda)  \frac{1 + 8 \frac{\lambda^{2}}{c^{2}}}{\lambda \left(1+ 4\frac{\lambda^{2}}{c^{2}}\right)} F(\lambda | \tilde{\lambda}^p_k , \tilde{\lambda}^h_k )\nonumber \\
  &\qquad + \log \left(\frac{\tilde{\lambda}^p_k   \sqrt{ ( \tilde{\lambda}^p_k/c)^2 + 1/4}}{\tilde{\lambda}_k^h   \sqrt{  ( \tilde{\lambda}_k^h/c)^2 + 1/4})  }\right) \Big]  \epp
\end{align}

\section{Asymptotics of the saddle point distribution}
The asymptotic expansion of the saddle point distribution $2\pi\rho(x)$ in Eq.~\eqref{eq:rho} for $x\to\infty$ and arbitrary fixed $0<\tau<\infty$ is given by (up to $\mathcal{O}\left(x^{-10}\right)$)
\begin{multline}
	2\pi\rho(x) \sim \frac{\tau^2}{x^4} - \frac{\tau^2-24\tau^3}{4x^6}
	+ \frac{\tau^2-120\tau^3+464\tau^4}{16x^8} \epp
\end{multline}
The first two terms were previously found within an approach using q-Bosons \cite{2013_Kormos_arXiv_1305}. The Tonks-Girardeau limit $c\to\infty$ at fixed density $n$, {\it i.e.}~$\tau\to 0$, can be easily performed,
\begin{align}
	a(x) &\sim a^{(0)}(x) = \frac{\tau^2}{x^2(x^2+1/4)} \notag\\
	\Rightarrow\quad 2\pi\rho(x) &\sim \frac{\tau^2}{x^2(x^2+1/4)+\tau^2} \epp
\end{align}
Substituting $\tau=n/c$ and $x=\lambda/c$ yields $2\pi\rhosp(\lambda) \sim 4n^2/(\lambda^2+4n^2)$. 

The limit $c\to 0$ at fixed density $n$, {\it i.e.}~$\tau\to\infty$, is more complicated. The first observation is that the scaled density $\rho_s(x)=2\rho(2\sqrt{\tau}x)/\sqrt{\tau}$, calculated by using the exact expression \eqref{eq:exacta}, has a proper limit, 
\begin{align}\label{eq:scaled_rho}
	\lim\limits_{\tau\to\infty}\rho_s(x) &= \lim\limits_{\tau\to\infty}\frac{2\rho(2\sqrt{\tau}x)}{\sqrt{\tau}} = \frac{2}{\pi}\sqrt{1-x^2}\ \Theta(1-x^2) \notag\\
	\Rightarrow\quad \rho(x) &\sim\frac{\sqrt{\tau}}{\pi}\sqrt{1-\frac{x^2}{4\tau}}\ \Theta(4\tau-x^2)\epp
\end{align}
This can be proven by analyzing the asymptotic behavior of Bessel functions. First of all we write 
\begin{align}\label{eq:two_factors_SupMat}
	2\pi\frac{2\rho(2\sqrt{\tau}x)}{\sqrt{\tau}} &= \frac{\sqrt{\tau}\partial_{\tau}a(2\sqrt{\tau}x,\tau)}{1+a(2\sqrt{\tau}x,\tau)} \nonumber \\
	& = \frac{a(2\sqrt{\tau}x,\tau)}{1+a(2\sqrt{\tau}x,\tau)}  \frac{\sqrt{\tau}\partial_{\tau}{a(2\sqrt{\tau}x,\tau)}}{a(2\sqrt{\tau}x,\tau)}  \epc
\end{align}
where the partial derivative is with respect to the second argument. We analyze the two factors separately. The first one becomes 
\begin{equation}\label{eq:sec_factor_rho}
	\lim\limits_{\tau\to\infty} \frac{a(2\sqrt{\tau}x,\tau)}{1+a(2\sqrt{\tau}x,\tau)} = \left\{\begin{array}{cl} 1 & \text{for}\quad |x|\leq 1\\ f(x) & \text{for}\quad |x|>1\end{array}\right.\epc
\end{equation}
where $f$ is a function with $f(1)=1$, and which decays algebraically as $(2x)^{-4}$ for large $x$. In order to analyze the second factor in Eq.~\eqref{eq:scaled_rho} we set $z=4\sqrt{\tau}$. Using formula \eqref{eq:exacta} and the abbreviation $\nu=xz$ we obtain, due to $\tau\partial_\tau = \frac{z}{2}\partial_z$ and $I'_\nu(z)=(I_{\nu+1}(z)+I_{\nu-1}(z))/2$, 
\begin{align}\label{eq:Re_quotient_Bessel}
	&\frac{\sqrt{\tau}\partial_{\tau}{a(2\sqrt{\tau}x,\tau)}}{a(2\sqrt{\tau}x,\tau)} \nonumber \\
	&= \frac{4}{z} + \frac{I_{- i\nu}(z)}{I_{1- i\nu}(z)} + \frac{I_{2- i\nu}(z)}{I_{1- i\nu}(z)}+ \frac{I_{ i\nu}(z)}{I_{1+ i\nu}(z)}+ \frac{I_{2+ i\nu}(z)}{I_{1+ i\nu}(z)}
	\nonumber \\
	&= 4\,\Re{\left[\frac{I_{- i\nu}(z)}{I_{1- i\nu}(z)}\right]}\epp
\end{align}
Note that the partial derivative acts only on the argument, but not on the order of the modified Bessel functions. Now we use the uniform asymptotic limit of the modified Bessel function of the first kind \cite{Dunster_1990}, which is also known as the uniform Airy-type asymptotic expansion of Bessel functions \cite{Gil_Comput_Mat}:
\begin{widetext}
\begin{multline}\label{eq:modBesselAsymp}
	I_{- i\nu}(\nu z') \sim \frac{  e^{\pi\nu/2}}{2\nu^{1/3}}\left(\frac{4\zeta}{1-{z'}^2}\right)^{1/4} \left\{   \left[\text{Bi}(-\nu^{2/3}\zeta)+2 i  e^{-\pi\nu}\sinh{(\pi\nu)}\text{Ai}(-\nu^{2/3}\zeta)\right] 
	\sum\limits_{s=0}^{\infty}(-1)^s\frac{A_s(\zeta)}{\nu^{2s}} \right. \\
\left. + \left[\text{Bi}'(-\nu^{2/3}\zeta)+2 i  e^{-\pi\nu}\sinh{(\pi\nu)}\text{Ai}'(-\nu^{2/3}\zeta)\right]\sum\limits_{s=0}^{\infty}(-1)^s\frac{B_s(\zeta)}{\nu^{2s+4/3}}     \right\}\epc
\end{multline}
\end{widetext}
where $\frac{2}{3}\zeta^{3/2} = \log{\left(\frac{1+\sqrt{1-{z'}^2}}{z'}\right)-\sqrt{1-{z'}^2}}$. $\text{Ai}$ and $\text{Bi}$ are Airy functions and the lowest expansion coefficients are given by $A_0=B_0=1$. Expanding the quotient in Eq.~\eqref{eq:Re_quotient_Bessel} for large $z$ to the leading asymptotic order we have to bear in mind that, in the denominator, imaginary order and argument are $\nu+ i$ and $z'\nu/(\nu+ i)$, respectively, instead of $\nu$ and $z'=1/x$ as in the numerator. Using the asymptotic expansions of Airy functions
\begin{equation}
	\text{Ai}(-z) \sim \frac{\sin{\left(\frac{2}{3}z^{3/2}+\frac{\pi}{4}\right)}}{\sqrt{\pi}z^{1/4}}\epc\:\:
	\text{Bi}(-z) \sim \frac{\cos{\left(\frac{2}{3}z^{3/2}+\frac{\pi}{4}\right)}}{\sqrt{\pi}z^{1/4}}\epc
\end{equation}
we eventually obtain by putting $z= \nu z' = \nu/x$ 
\begin{multline}
	\Re{\left[\frac{I_{- i xz}(z)}{I_{1- i xz}(z)}\right]_{z\to\infty}}
	 = \Re{\left[\frac{I_{- i\nu}(\nu z')}{I_{1- i\nu}(\nu z')}\right]_{\substack{\nu=xz \to\infty\\ z'=1/x<\infty}} } 
	\\
	\sim \Re{\left[\frac{1+\sqrt{1- {z'}^2}}{i z'}\right]_{\substack{\nu=xz \to\infty\\ z'=1/x<\infty}} }
	= \Theta(1-x^2)\sqrt{1 - x^2}\epp
\end{multline}
In the last step we used that, due to the real part, the leading order is only non-zero if the absolute value of $x$ is less than one. For $|x|>1$ the leading order vanishes and we have $\lim_{\tau\to\infty}\rho_s(x) = 0$ for every fixed $|x|>1$. Taking the factors $2\pi$ and $4$ in Eqs.~\eqref{eq:two_factors_SupMat} and \eqref{eq:Re_quotient_Bessel} into account we finally get $\rho_s(x) = \frac{2}{\pi}\sqrt{1 - x^2}$ for $|x|\leq 1$ and $\rho_s(x) = 0$ elsewhere, which proves Eq.~\eqref{eq:scaled_rho}. 

Substituting $\tau=n/c=1/\gamma$, $x=\lambda/c$ in Eq.~\eqref{eq:scaled_rho} and defining $\lambda_\ast=2n\sqrt{\gamma}$, we obtain
\begin{equation}
	\rhosp(\lambda) \sim n\frac{2}{\pi\lambda_\ast}\sqrt{1-\frac{\lambda^2}{\lambda_\ast^2}}\ \Theta(\lambda_\ast-|\lambda|) \epp
\end{equation}
which reproduces the leading term of the ground state distribution of the Lieb-Liniger model in the low-$\gamma$ expansion \cite{2002_Wadati}.

\section{Time evolution of the density-density correlations for $\gamma = \infty$}\label{sec:app_time_evo}
In order to determine the time evolution of the density-density correlator, on needs the density-density matrix-elements between two different $N$-particle Bethe states $|\lambda \rangle$ and $| \bar{\lambda}\rangle$ in the limit $c \to \infty \epp$  Using the first quantized version of the density operator $\hat{\rho}(x) = \sum_{j=1}^N \delta(x - x_j)$ and the standard expression for the wave function for the Tonks-Girardeau gas, we have
\begin{align}
\langle \bar{\lambda} & |\ :\hat{\rho}(x)\hat{\rho}(0) : \, |\lambda \rangle\ \nonumber \\ = &
\  \frac{1}{N!\, L^{N}} \int_{0}^{L}\mathrm{d}^{N}x\, 
\left( \sum_{P}(-1)^{[P]}\prod_{j=1}^{N}   e^{i\,x_{j}\lambda_{P_{j}}} \right) \nonumber\\ \:\: & \times
\left( \sum_{P'}(-1)^{[P']}\prod_{k=1}^{N}   e^{-i\,x_{k}\bar{\lambda}_{P'_{k}}} \right)\sum_{l,m=1}^{N}\delta(x-x_{l})\,\delta(x_{m}) \nonumber \\ = &
\  \frac{1}{(N-2)!\, L^{N}} \sum_{P,P'}(-1)^{[P]+[P']}  e^{i\, x (\lambda_{P_{1}}-\bar{\lambda}_{P'_{1}})}\nonumber \\ \:\: &
\times \left( \prod_{j=3}^{N} \int_{0}^{L}\mathrm{d}x_{j}\,   e^{i\,x_{j}(\lambda_{P_{j}}-\bar{\lambda}_{P'_{j}})} \right) \epp
\end{align} 
The product of $N-3$ integrations is only non-zero if the two states are the same up to at most two rapidities. This is a well-known property of the density operator acting on Bethe states at $c = \infty$. Since both the left and right state have to be parity invariant we focus on states which differ only by two rapidities:
\begin{equation}
\bar{\lambda}_{j}\ =\ \lambda_{j} \qquad \text{for} \quad j=3,4,\dots,N \epc
\end{equation}
and two rapidities $\lambda_{1}$ and $\lambda_{2}$ are, in general, different from $\bar{\lambda}_{1}$ and $\bar{\lambda}_{2}$. The integrals are non-zero only if $P_{j}=P'_{j}$ for $j=3,4,\dots,N$ and if the permutation $P\in S^{N}$ is restricted by
\begin{equation}
P_{j}\in \{3,4,5,\dots,N\} \qquad \text{for} \quad j=3,4,5,\dots,N \epp
\end{equation}
There are $2\cdot(N-2)!$ such permutations, namely $(N-2)!$ permutations for which $P_{1}=1$ and $P_{2}=2,$ and another $(N-2)!$ permutations for which $P_{1}=2$ and $P_{2}=1.$ The permutation $P'$ is almost completely fixed by $P,$ the only possible choices are
\begin{align}
&\left( P'_{1}=P_{1}\quad \text{and} \quad P'_{2}=P_{2}\right)  \notag\\[1ex] 
\text{or}\quad  &\left( P'_{1}=P_{2}\quad \text{and} \quad P'_{2}=P_{1}\right) \epp
\end{align}
We thus obtain
\begin{align}
&\frac{1}{(N-2)!\, L^{N}} \sum_{P,P'}(-1)^{[P]+[P']} e^{i\, x (\lambda_{P_{1}}-\bar{\lambda}_{P'_{1}})} \nonumber \\ \:\: & \times 
\left( \prod_{j=3}^{N} \int_{0}^{L}\mathrm{d}x_{j}\,   e^{i\,x_{j}(\lambda_{P_{j}}-\bar{\lambda}_{P'_{j}})} \right)   
 \nonumber \\=&
\  \frac{1}{L^{2}}  \sum_{P_{\text{truncated}}} \left(   e^{i\, x (\lambda_{P_{1}} - \bar{\lambda}_{P_{1}})} -   e^{i\, x (\lambda_{P_{1}} - \bar{\lambda}_{P_{2}})} \right) \nonumber \\=  & \
\frac{1}{L^{2}}  \Big(   e^{i\, x \lambda_{1} } -    e^{i\, x \lambda_{2} } \Big)\Big(  e^{-i\, x\bar{\lambda}_{1}} -   e^{-i\, x \bar{\lambda}_{2} } \Big) \epp
\end{align}
With this expression for off-diagonal matrix elements, we can now recover the whole time evolution of the density-density operator. The only contribution to the sum
\begin{multline}
\langle 0 | e^{i H_{LL}t }\hat{\rho}(x)\hat{\rho}(0)  e^{-i H_{LL}t }| 0 \rangle \\
=
\frac{1}{2} \sum_{ \mathbf{e} }   e^{ - \delta s_\mathbf{e} -  i \delta\omega_\mathbf{e} t } \langle \rhosp,\emptyset | \hat{\rho}(x)\hat{\rho}(0)  | \rhosp , \mathbf{e} \rangle 
\\
+ \frac{1}{2} \sum_{ \mathbf{e} }    e^{ - \delta s^\ast_\mathbf{e} +  i \delta\omega_\mathbf{e} t } \langle \rhosp,\mathbf{e} | \hat{\rho}(x)\hat{\rho}(0) | \rhosp , \emptyset \rangle 
\end{multline}
comes from particle-hole excitations {consisting of only one parity-invariant pair}: $$\{\bar{\lambda}_1 =  \lambda_p,\bar{\lambda}_2 =  - \lambda_p| \lambda_1 = \lambda_h, \lambda_2 = - \lambda_h \} .$$ 
Their spectrum is the one of free particles $\delta\omega_{\mathbf{e}} = 2 \lambda_p^2 - 2 \lambda_h^2$ and since the shift function \eqref{eq:shift_F} is trivial, the difference of the two overlap coefficients is simply $\delta s_{\mathbf{e}} = -\log \lambda_p + \log \lambda_h$. Putting everything together we obtain
\begin{align}
\langle 0 | & e^{i H_{LL}t }\hat{\rho}(x)\hat{\rho}(0)  e^{-i H_{LL}t }| 0 \rangle
- \langle\hat{\rho}(x) \hat{\rho}(0) \rangle_{\text{sp}} \notag\\[1ex]
 & = \sum_{\lambda_p>0 } \sum_{\lambda_h>0}   e^{ - \delta s_\mathbf{e} -  i \delta\omega_\mathbf{e} t } \Big[\frac{4}{L^2} \sin{(\lambda_h x)}\sin{(\lambda_p x)} \Big] \notag\\ 
 &\to\left|\int_{-\infty}^\infty \frac{dk}{ \pi}  \frac{n k}{4 n^2 + k^2 }   e^{- 2it k^2  + i k x }  \right|^2 \epc
\end{align}
where we used the saddle point distribution in the fermionized regime $\rhosp(\lambda) = \frac{1}{2 \pi} \frac{1}{(\lambda/2n)^2 +1}$ and $\rho^h(\lambda) = \frac{1}{2 \pi} - \rhosp(\lambda)$.

\end{document}